# FLEX: FLEXible Federated Learning Framework

**Francisco Herrera** [a], **Daniel Jiménez-López** [a], **Alberto Argente-Garrido** [a],

**Nuria Rodríguez-Barroso** [*,a], **Cristina Zuheros** [a], **Ignacio Aguilera-Martos** [a],

**Beatriz Bello** [a], **Mario García-Márquez** [a], **M. Victoria Luzón** [b]

[a] *Department of Computer Science and Artificial Intelligence, Andalusian Research Institute in Data Science and Computational Intelligence (DaSCI), University of Granada, Spain*
[b] *Department of Software Engineering, Andalusian Research Institute in Data Science and Computational Intelligence (DaSCI), University of Granada, Spain*

## Abstract

In the realm of Artificial Intelligence (AI), the need for privacy and security in data processing has become paramount. As AI applications continue to expand, the collection and handling of sensitive data raise concerns about individual privacy protection. Federated Learning (FL) emerges as a promising solution to address these challenges by enabling decentralized model training on local devices, thus preserving data privacy. This paper introduces FLEX: a FLEXible Federated Learning Framework designed to provide maximum flexibility in FL research experiments. By offering customizable features for data distribution, privacy parameters, and communication strategies, FLEX empowers researchers to innovate and develop novel FL techniques. The framework also includes libraries for specific FL implementations including: (1) anomalies, (2) blockchain, (3) adversarial attacks and defences, (4) natural language processing and (5) decision trees, enhancing its versatility and applicability in various domains. Overall, FLEX represents a significant advancement in FL research, facilitating the development of robust and efficient FL applications.

* Corresponding Author
  Email addresses: `herrera@decsai.ugr.es` (Francisco Herrera) `aargente@ugr.es` (Alberto Argente-Garrido), `dajilo@ugr.es` (Daniel Jiménez-López), `rbnuria@ugr.es` (Nuria Rodríguez-Barroso), `czuheros@ugr.es` (Cristina Zuheros), `nacheteam@ugr.es` (Ignacio Aguilera), `bbgarcia@ugr.es` (Beatriz Bello), `mariogmarq@correo.ugr.es` (Mario García-Márquez), `luzon@ugr.es` (M. Victoria Luzón).

# 1 Introduction

In the realm of Artificial Intelligence (AI), the extensive use of data presents significant challenges regarding privacy and security. As AI applications broaden to address various issues, the collection and processing of sensitive data intensify, raising concerns over individual privacy protection. Furthermore, efficient and secure communication of this data between devices and central servers becomes crucial to prevent potential vulnerabilities. Balancing the utility of AI with the safeguarding of privacy becomes an ethical imperative, driving the search for innovative approaches that address these challenges comprehensively [1].

In this context, Federated Learning (FL) [2] emerges as a promising solution to address privacy and communication concerns in the AI era. Rather than centralizing data on a single server, FL enables model training to occur in a decentralized manner on local devices. Only aggregated updates are shared with a central server, thus preserving the privacy of individual data. This methodology not only addresses privacy concerns but also promotes efficient collaboration in model improvement, utilizing the diversity of distributed data. FL represents an innovative and ethical approach to leveraging the power of AI while protecting the rights and security of users [3].

When researching in FL, simulating it is neccesary as it allows for thorough testing of privacy measures, algorithm refinement, performance evaluation, scalability analysis, robustness assessment, resource planning, and the exploration of novel techniques, while mitigating risks associated with handling sensitive data and multiple device connections in real-world scenarios [4]. Nevertheless, this experimentation is not trivial due to the intricate interplay of decentralized data, privacy-preserving mechanisms, device heterogeneity, non-IID data, dynamic aspects, and the need for meaningful benchmarks [5]. Overcoming these challenges is crucial to advancing research in FL and ensuring that FL algorithms are robust and effective in practical, real-world scenarios. While existing FL frameworks address these problems individually, to the best of our knowledge, none of them provide enough flexibility to model each aspect of a FL experiment.

A novel research FL framework that offers maximum flexibility is crucial. Researchers should be able to customize and experiment with various aspects of FL simulations, such as data distribution, privacy parameters, and communication strategies. This flexibility empowers researchers to explore novel FL techniques, fine-tune existing ones, and adapt FL to specific use cases without the burden of dealing with the intricacies of simulation boilerplate code. In essence, simulated FL frameworks play a pivotal role in accelerating research in FL by providing a controlled environment that mirrors real-world complexities while allowing researchers the freedom to innovate and develop novel FL techniques.

To deal with each one of the complexities of researching in FL while allowing maximum flexibility, we present *FLEX: a FLEXible Federated Learning Framework*, an open-source library with license Apache 2.0 written in Python. FLEX implements mechanisms to model almost every non-IID data distribution, flexibility to design custom FL architectures as well as a hihgly customizable FL pipeline with multiple and interchangeable granularity levels of abstraction. It provides transparent support for popular deep learning and machine learning frameworks such as Pytorch, Tensorflow, Hugginface libraries or Scikit-Learn. Such support is not limited to their models, it also supports simulating a federated distribution of their builtin datasets. Furthermore, FLEX is published along with several companion open-source libraries to bring specific FL datasets and implementations from the fields of anomaly detection, blockchain, adversarial attacks and defences, Natural Language Processing (NLP) and decision trees, namely: *FLEX-Anomalies, FLEX-Block FLEX-Clash, FLEX-NLP and FLEX-Trees*.

According to these aims, the paper is organized as follows. Section 2 introduces and motivates FLEX by addressing the main challenges in FL research experiments. Section 3 instructs the user on how to install FLEX and describes the quality standards of the code developing process. Then, Section 4 presents the libraries built on top of FLEX, followed by an extensive comparison with existing FL frameworks in Section 5. Finally, Section 6 summarises the conclusions of this paper and future work.

## 2 FLEX Framework Overview

We present the main challenges faced when simulating FL to then introduce how FLEX excels at fulfilling them while preserving maximum flexibility in Section 2.1. Finally, we provide some software description and usage as short code examples in 2.2.

### 2.1 Challenges in FL research experiments

Researching in FL is a challenging task due to the inherent complexities and unique characteristics of FL that distinguish it from traditional centralized ML. Several factors contribute to the non-trivial nature of FL simulation, in the following we discuss the most relevant ones:

- *Decentralized Data Sources.* In FL, data is distributed across multiple devices, servers, or edge nodes. Simulating this decentralized data distribution accurately requires modeling diverse data sources with varying data sizes, types, and distributions. Furthermore, FL can be categorized based on the dimension in which the data is partitioned across clients [2] in Horizontal FL (HFL), Vertical FL (VFL) and Federated Transfer Learning (FTL). The challenges among different FL categorizations also vary, for example, the main challenge in HFL is simulating non-Independent and Identically Distributed (non-IID) data and the main challenge in VFL is simulating realistic splits of features among FL nodes. Thus, the challenge of simulating decentralized data sources is multiple and depends on the FL categorization.

- *Accounting for the actual distribution of heterogeneous FL nodes.* FL often involves devices with varying computational capabilities, memory constraints, communica-tion bandwidths and geographical locations. Accurately modeling these heterogeneous devices and their impact on the learning process is essential for realistic simulations. Such heterogeneity leads to multiple organizations of FL nodes or architectures, modelled after the capabilities of the nodes. Furthermore, it must be considered that each FL category does not necessarily implement the same architecture. Thus, to model the distribution of FL nodes, a framework to simulate FL faces the challenge of bringing support to as many FL architectures as possible.

- *Customized FL flows and support for popular ML frameworks.* FL encompasses complex algorithms for federated aggregation, model updates, and convergence. Simulated FL frameworks should be flexible enough to allow exploring novel techniques and variations of FL, while providing reference implementations or benchmarks, crucial for understanding the strengths and limitations of different FL approaches and making informed decisions about their suitability for specific applications. Many novel ML models are highly dependent on a specific ML framework, even though there might exist frameworks with similar features, and porting it to another framework can be very time-consuming, thus providing a ML framework-agnostic simulated FL framework is ideal. Enabling to simulate and implement the final FL model in the same ML framework. Thus, a simulated FL framework must provide enough granularity for extending its capabilities without rewriting its source code and transparent support for popular ML frameworks.

To summarize, a simulated FL framework should be able to model the inherent complexities of a FL model, while providing enough flexibility to develop and test novel FL techniques using any popular centralized ML framework.

## 2.2 Software description and usage

FLEX is designed as a set of decoupled modules, that maximize the ability to implement novel FL algorithms, without the burden of the repetitive implementation of basic FL settings, such as federated data partition or federated architecture design and implementation. Thus, we have split the design of FLEX into three main modules represented in Figure 1: the data module (see Section 2.2.1), the actor module (see Section 2.2.2) and the pool module (see Section 2.2.3).

### 2.2.1 Data module

As the decentralized data sources are a key challenge to simulating FL, we have designed an entire module, `flex.data`, to handle the partition of data among nodes, as well as, to implement the main datasets of the FL literature acknowledged in [6], such as Sentiment140, FederatedMNIST, CelebA and Shakespeare.

To illustrate the capabilities of this module, we devise two use cases:

- Use Case 1 (UC1): Simulate a HFL data split using the CIFAR10 dataset from Pytorch that fits the following description from [7]: *"sample two/ten classes for each*

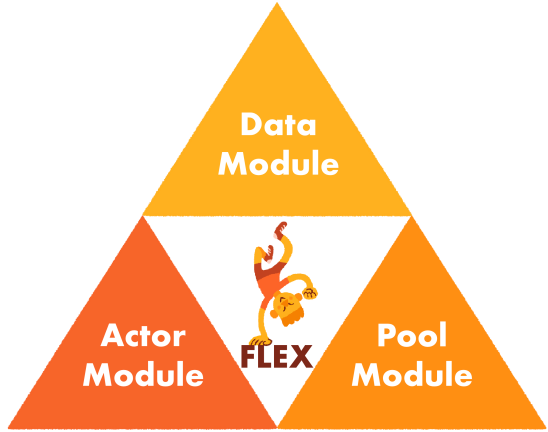

Figure 1: Representation of the module design of FLEX.

*client for CIFAR10/CIFAR100; Next, for each client i and selected class c, we sample $\alpha_{i,c} \sim U(.4, .6)$, and assign it with $\frac{\alpha_{i,c}}{\sum_j \alpha_{j,c}}$ of the samples for this class. We repeat the above using 10, 50 and 100 clients."*

- Use Case 2 (UC2): Simulate a FTL data split using the Kaggle's Default-of-CreditCard-Clients dataset that fits the following description from [8]: *"split the dataset [...] to simulate a two-party federation problem. Specifically, we assign each sample to party A, party B or both so that there exists a small number of samples overlapping between A and B. All labels are on the side of party A. We will examine the scalability (see the "Scalability" section) of the FTL algorithm by dynamically splitting the feature space."* In the Scalability section, they make each client have from 5 to 30 features.

The implementation in FLEX of each use case is provided in code listings 1 and 2, for simplicity, we omit boilerplate code required for loading the datasets into `cifar10` and (`defaul_creditcard_x_data`, `defaul_creditcard_y_data`), respectively. The general workflow to simulate a federated data partition with FLEX is as shown in Figure 4: create a `flex.data.Dataset` object, which is designed to provide a transparent layer to import and export datasets from other popular dataset libraries such as HugginFace datasets, tensorflow_datasets and torchvision. Then, define a `flex.data.FedDatasetConfig` object that fits the requirements of the simulated data partition and finally, provides both, the `flex.data.Dataset` and the `flex.data.FedDatasetConfig` objects to the class method `FedDataDistribution.from_config`, which performs the partition and returns a dictionary where each key is an id for the data partition, paired with its value a `flex.data.Dataset` object, that represents an individual split.

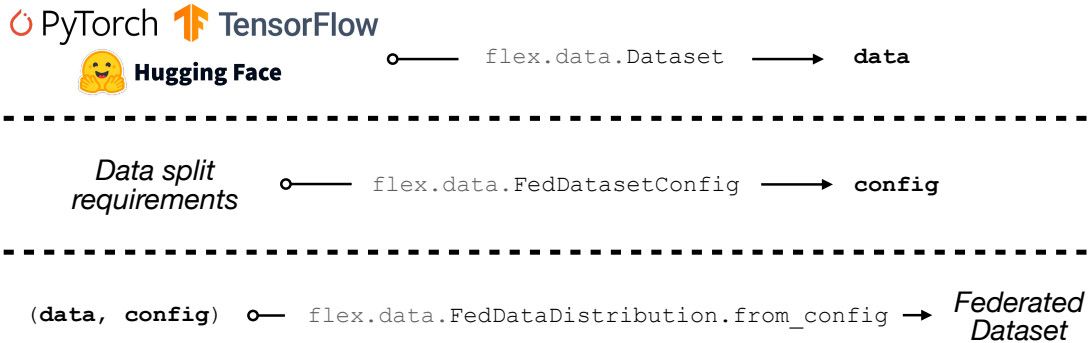

Figure 2: Workflow to simulate a federated data partition with FLEX.

```
1  train_dataset = flex.data.Dataset.from_torchvision_dataset(cifar10)
2  # Fix a seed to make our split reproducible
3  config = flex.data.FedDatasetConfig(seed = 0)
4  # it is not clear whether clients share their data or not
5  config.replacement = True
6  # 10 nodes, for greater number of nodes, change this value. As we do not provide a
   ↪ custom id for each node, integers starting from 0 are used as id
7  config.n_nodes = 10
8  # Assign a sample proportion for each node-class pair
9  num_classes = 10
10 alphas = numpy.random.uniform(0.4, 0.6, [config.n_nodes, num_classes])
11 alphas = alphas / numpy.sum(alphas, axis=0)
12 config.weights_per_class = alphas
13 # Perform the actual data split
14 federated_dataset = flex.data.FedDataDistribution.from_config(train_dataset, config)
```

Listing 1: Implementation of UC1 in FLEX.

### 2.2.2 Actor module

This module is responsible for creating the architecture of a simulated FL scenario. We define three main roles of each FL node and we define an actor as a node with a set of roles, that is, an actor can have multiple roles assigned. The set of roles assigned to an actor is a hint to indicate whether it begins the communication path with other actors and is implemented in class `flex.actors.FlexRole`. Note that, once the communication is established, nodes can communicate bidirectionally:

- *Client role:* this role is the most restrictive, as it does not allow establishing communications with other nodes.

- *Aggregator role*: it allows communicating with other nodes with the same role or with the server role.

```
1   train_dataset = flex.data.Dataset.from_numpy(defaul_creditcard_x_data,
    ↪ defaul_creditcard_y_data)
2   # Fix a seed to make our split reproducible
3   config_ftl = flex.data.FedDatasetConfig(seed = 0)
4   # Two nodes A and B
5   config_ftl.n_nodes = 2
6   # Provide a custom id for each node. Otherwise, numbers starting from 0 are used as id
7   config_ftl.node_ids = ["Node A", "Node B"]
8   # Some samples are shared
9   config_ftl.replacement = True
10  # By specifying weights, we ensure that nodes do not share every sample, each node owns
    ↪ 75% of the samples with some overlapping
11  config_ftl.weights = [0.75, 0.75]
12  # Each node owns 5 features, change this number to increase the number of features per
    ↪ node
13  feat_per_node = 5
14  config_ftl.features_per_node = [feat_per_node, feat_per_node]
15  # Node A keeps all labels, while Node B has none
16  config_ftl.keep_labels = [True, False]
17  # Perform the actual data split
18  federated_dataset = flex.data.FedDataDistribution.from_config(train_dataset,
    ↪ config_ftl)
```

Listing 2: Implementation of UC2 in FLEX.

- *Server role*: it allows communicating with other nodes with the same role or with the client role.

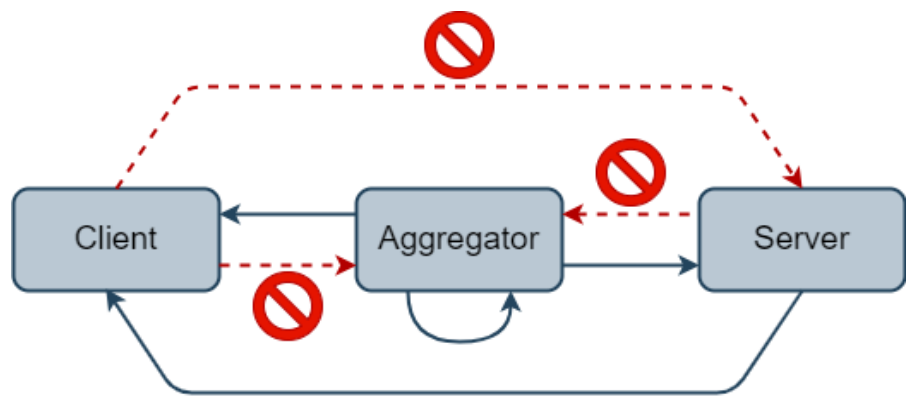

Figure 3: Visual description of the permissions associated with the Client, Aggregator and Server roles.

Figure 3 illustrates the communication path of each role. We stress that an actor can have multiple roles, fixed in the creation of the architecture and cannot change during the FL simulation.

FL architectures in FLEX are designed to be a set of actors with one or multiple roles. The set of actors and their roles is implemented as a dictionary in `flex.actors.FlexActors`, where each key is an actor id and its value is a `flex.actors.FlexRole`. Each of the following FL architectures can be implemented in FLEX, as shown in code listing 3:

- Client-server: of N actors, one has the server and aggregator roles while the rest only has the client role. This is the most usual FL architecture in HFL simulations.

- Peer-to-peer: every actor has the three roles, client, server and aggregator. There is no central authority, so that, each actor can aggregate and distribute updated models to every actor. This architecture is well suited for VFL, as each actor participates in the aggregation step.

Note that, we assume that `actor_ids` is a list with the ids of each client actor.

```python
# Client-server architecture
actors = FlexActors()
for aid in actor_ids:
    actors[aid] = FlexRole.client
# Custom id for server node
actors["server"] = FlexRole.server
# Alternatively, this architecture is built in FLEX
actors = flex.actors.client_server_architecture(clients_ids = actor_ids, server_id =
    "server")

# Peer-to-peer architecture
actors = FlexActors()
# Every node has every role
for aid in actor_ids:
    actors[aid] = FlexRole.server_aggregator_client
# Alternatively, this architecture is built in FLEX
actors = flex.actors.p2p_architecture(nodes_ids = actor_ids)
```

Listing 3: Implementation of multiple FL architectures in FLEX.

### 2.2.3 Pool module

Once we have modelled FL data splits and architectures, we need to model the FL flow and the model of each actor. Given that the implementation of each ML model varies, we represent a general ML model with a `FlexModel` object, which is a dictionary with an additional property, the id of the model owner, `actor_id`. This implementation allows FLEX to be model-agnostic, given that we rely on the key-value storage of the dictionary to store and access the implementation-specific entities of each ML model. Then, it is possible to define a pool of communications, a pool for short, responsible for implementing the communication dynamics. A pool is a set of actors, data and models, each one of them indexed using the id of each actor so that each actor has data, a set of roles and a model. It is implemented in

`FlexPool`. The communication dynamics of an FL flow are modelled after the interaction of `FlexPool` objects using `select` and `map` methods.

The `select` method allows retrieving a smaller pool from a bigger one using the id and role of each actor as criterion. For example, it allows extracting a pool with actors having the client role or selecting actors based on their id as shown in code listing 4, where we assume `pool` to be a `FlexPool` with one server-aggregator actor and three client actors with ids "0", "1" and "2".

```python
# Pool with two random actors
subpool = pool.select(2)
# Pool with actors with server role
servers = pool.servers
# alternatively
servers = pool.select(lambda actor_id, set_of_roles:
    FlexRoleManager.is_server(set_of_roles))
# Pool with actors with ids 0 and 2
selected_clients = pool.select(lambda actor_id, set_of_roles: actor_id in ["0", "2"])
```

Listing 4: Example of using `select` to extract smaller pools in FLEX.

The `map` method is designed to establish communication among two pools, source and destination, so that, the source pool can access a read-only version of the `FlexModel`'s of the destination pool. These low granularity communication tools permit modelling complex FL flows. FLEX builds two higher level abstractions to provide more user-friendly usage, namely decorators and primitives. Decorators try to hide the low-level complexities of FL communication flows while being framework-agnostic, and primitives go a step further and provide ML framework specific implementations for FL flows. All of them can be used interchangeably, allowing maximum code reusability while preserving flexibility. To illustrate these features, we model the HFL step in which the weights of client's models are collected by the server model in code listing 5, employing Tensorflow. First, we show how it can be implemented using map, then with decorators and finally Tensorflow primitives. We omitted boilerplate code, assumed that a Tensorflow model is stored in each client's `FlexModel` using key `model` and collected client's weights in server's `FlexModel` using key `weights`.

For further examples of usage of the library, we refer to the extensive collection of well documented Jupyter notebooks, that showcase how to use sklearn, Tensorflow and Pytorch with FLEX leveraging the described features.

## 3  Installation and design quality standards

The FLEX framework can be installed using PyPi using `pip install flex`. It is also available by cloning the repository from GitHub[1] and executing, the command `pip install .`

---
[1] https://github.com/FLEXible-FL/FLEXible

```
1   # Lowest level
2   def get_clients_weights(server_flex_model: FlexModel, clients_flex_models:
    ↪ list[FlexModel]):
3       server_flex_model["weights"] = []
4       for k in clients_flex_models:
5           client_weights = clients_flex_models[k]["model"].get_weights()
6           server_flex_model["weights"].append(client_weights)
7
8   # With decorators from FLEX
9   from flex.pool import collect_clients_weights
10  # The decorated function must return the weights of the model
11  @collect_clients_weights
12  def get_clients_weights(client_flex_model: FlexModel):
13      return client_flex_model["model"].get_weights()
14
15  # With Tensorflow primitives from FLEX
16  from flex.pool import collect_clients_weights_tf as get_clients_weights
17
18  # In either case, all of them are used as follows:
19  servers.map(get_clients_weights, selected_clients)
```

Listing 5: Example of using map showing three possible levels of granularity, namely, low-level, decorators and primitives using a Tensorflow model in FLEX.

in the root directory. After that, the package will be available for its usage within the name flex. An extensive documentation is provided, hosted on the Read the Docs[2] platform.

We highlight that each FLEX module is the result of a thoughtful study from both the software and the FL perspective, to choose the software design pattern that best suited its functionality following SOLID principles [9]. The software design patterns implemented for each module are:

- Data module: The software pattern chosen was Builder or Constructor.

- Actor module: The software pattern chosen was Mediator.

- Pool module: The software pattern chosen was Mediator. In addition, each level of abstraction in its usage has its own pattern; low-level retains the Mediator pattern, decorators use the pattern Decorator and primitives use the pattern Facade.

With the description of the software patterns employed and the functionality of each module, we are welcoming the community to improve and extend the design of FLEX as well as FLEX companion libraries. Moreover, the entire framework's functionality is covered by an extensive battery of tests that ensure its proper operation and facilitate the incorporation of new features, while minimizing the occurrence of bugs and unexpected behaviors.

---

[2]https://flexible.readthedocs.io/en/latest/

## 4 FLEX companion libraries

FLEX serves as a foundational framework that empowers developers and researchers to simulate and create FL adaptations of prominent ML disciplines. To complement this core framework, we have developed a collection of libraries that extend the capabilities of FLEX. In the following section, we will delve into the details of these companion libraries, shedding light on how they facilitate and enrich the practice of FL across different domains, from decision trees to anomaly detection.

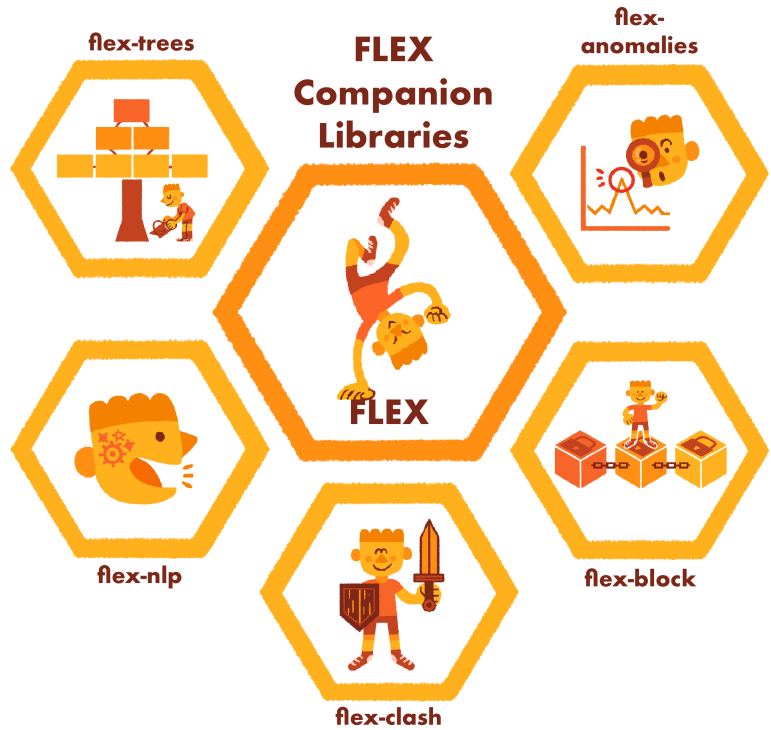

Figure 4: Representation of all the companion libraries implemented to complement FLEX.

**FLEX-Anomalies** Anomaly detection represents a significant area of focus within the field of ML [10]. This discipline consists on evaluating the deviation or abnormality present in the data. It serves dual purposes: firstly, the purification and refinement of datasets, and secondly, identifying these anomalies for more in-depth analysis. This library includes a representative sample of algorithms and techniques related to anomaly detection, among which we highlight:

- Anomaly detection models based on distance, density and trees for static data [11].

- Neural network models for both time series and static data including a convolutional and recurrent architecture [12], alongside a fully connected AutoEncoder [13].

- Aggregators for all implemented models.
- Anomaly score processing techniques both federated and at the central server.
- Additional pre-processing techniques to deal with the data.

This library is present in GitHub[3] and the PyPi repository, available for installation under the command `pip install flex-anomalies`. After that, the package will be ready to use within the name `flex-anomalies`.

**FLEX-Block**  Introducing blockchain technology into FL has been proven useful since it enables a decentralized FL environment without a single point of failure and improved scalability [14]. This library implements both:

- Proof of Federated Learning [15], Proof of Stake and Proof of Work architectures.
- An interface for creating a custom blockchain architecture and using it in a FLEX program.

This library is present in in GitHub[4] and the PyPi repository, available for installation under the command `pip install flex-block`. After that, the package will be ready to use within the name `flex-block`.

**FLEX-Clash**  Defending against adversarial attacks in FL is paramount [16] because it ensures the privacy, security, and integrity of the learning process across distributed networks. By safeguarding against these threats, we can maintain the trustworthiness of the FL process, protect user privacy, and ensure that the collaborative benefits of FL are realized without compromising security. For that reason, this library integrates the state-of-the-art of both:

- Poisoning adversarial attacks [17] including from data to model poisoning and from byzantine to backdoor [18] attacks.
- Defense mechanisms against adversarial attacks.

This library is present in in GitHub[5] and the PyPi repository, available for installation under the command `pip install flex-clash`. After that, the package will be available for its usage within the name `flex-clash`.

**FLEX-NLP**  This library integrates widely used datasets and specific implementations to solve different natural language processing tasks:

---

[3] https://github.com/FLEXible-FL/flex-anomalies
[4] https://github.com/FLEXible-FL/flex-block
[5] https://github.com/FLEXible-FL/flex-clash

- Question Answering (QA). We integrate the Stanford Question Answering Dataset[6] (SQuAD), a widely used benchmark to advance the development of QA systems. We adapt a distilled version of the BERT base model, DistilBERT [19]. We use the clipped average aggregation, which includes an additional step of clipping the model updates before averaging,

- Semantic Textual Similarity (STS). We use the QQP-Tripets dataset, an adaptation of Quora Question Pairs[7] (QQP) dataset where data are organized as triplets: original query, similar query and non-similar query. We adapt a distilled version of the RoBERTa base model [20]. We use the weighted average aggregation where weights are set randomly, and

- Sentiment Analysis (SA). We consider the Internet Movie DataBase[8] (IMDB) for binary sentiment classification. We have designed a basic neural network from scratch to offer more variety in the design and adaptation of deep learning models to the federated environment [21]. We use the federated average aggregator.

This library is present in in GitHub[9] and the PyPi repository, available for installation under the command `pip install flex-nlp`. After that, the package will be available for its usage within the name `flex-nlp`.

**FLEX-Trees** This library integrates widely used datasets and specific implementations of state-of-the-art decision tree models for FL. The models implented cover both single and ensemble models, and those are:

- Federated ID3 (FedID3)[22]: Implements the classical ID3 tree in the federated environment,

- Federated Random Forest (FedRF)[23]: The classical Random Forest trained in the federated environment, and

- Federated Gradient Boosting Decision Tree (FedGBDT)[24]: The GBDT algorithm adapted to the federated environment.

The datasets added to this library are tabular datasets. These datasets are: Nursery [10], Adult [11], Car [12], Bank [13], ILDP [14], Credit2 [15] and Magic [16].

---

[6]https://rajpurkar.github.io/SQuAD-explorer/
[7]https://quoradata.quora.com/First-Quora-Dataset-Release-Question-Pairs
[8]https://ai.stanford.edu/ amaas/data/sentiment/
[9]`https://github.com/FLEXible-FL/flex-nlp`
[10]`http://archive.ics.uci.edu/dataset/76/nursery`
[11]`https://archive.ics.uci.edu/ml/datasets/adult`
[12]`http://archive.ics.uci.edu/dataset/19/car+evaluation`
[13]`http://archive.ics.uci.edu/dataset/222/bank+marketing`
[14]`https://archive.ics.uci.edu/dataset/225/ilpd+indian+liver+patient+dataset`
[15]`https://archive.ics.uci.edu/ml/datasets/default+of+credit+card+clients`
[16]`http://archive.ics.uci.edu/dataset/159/magic+gamma+telescope`

This library is present in in GitHub[17] and the PyPi repository, available for installation under the command `pip install flex-trees`. After that, the package will be available for its usage within the name `flex-trees`.

## 5 Comparison with other FL frameworks

This section is devoted to showcasing how FLEX compares to existing simulated FL frameworks, in terms of the above-described challenges in simulating FL, namely: decentralized data sources, the distribution of heterogeneous FL nodes and customized FL flows. Furthermore, we added a new subsection to account for desirable properties of FL frameworks from the perspective of software engineering, such as support for popular ML frameworks or documentation and tutorials.

Table 1 shows multiple FL frameworks and whether they fulfill those important aspects of simulated FL. Due to space limitations because of the large number of frameworks, the names of the different frameworks have been shrunk in the table. The frameworks reviewed are: PySyft (PyS)[18], Tensorflow Federated (TFF)[19], FATE (FAT)[20], PaddleFL (Pad)[21], Flower (Flo)[22], IBM Federated Learning (IBM)[23], Substra (Sub)[24], OpenFL (OFL)[25], FedML-AI (FML)[26], FedJAX (FJx)[27], FedLab (FLb)[28], SimFL (SFL)[29], EasyFL (EFL)[30], TorchFL (TFL)[31], APPFL: Argonne Privacy-Preserving Federated Learning (AFL)[32] and NVFlare (NVF)[33].

---

[17]https://github.com/FLEXible-FL/flex-trees
[18]https://github.com/OpenMined/PySyft [Accessed on February 19th, 2024]
[19]https://github.com/tensorflow/federated [Accessed on February 19th, 2024]
[20]https://github.com/FederatedAI/FATE [Accessed on February 19th, 2024]
[21]https://github.com/PaddlePaddle/PaddleFL [Accessed on February 19th, 2024]
[22]https://github.com/adap/flower [Accessed on February 19th, 2024]
[23]https://github.com/IBM/federated-learning-lib [Accessed on February 19th, 2024]
[24]https://github.com/Substra/substra [Accessed on February 19th, 2024]
[25]https://github.com/intel/openfl [Accessed on February 19th, 2024]
[26]https://github.com/FedML-AI/FedML [Accessed on February 19th, 2024]
[27]https://github.com/google/fedjax [Accessed on February 19th, 2024]
[28]https://github.com/SMILELab-FL/FedLab [Accessed on February 19th, 2024]
[29]https://github.com/Xtra-Computing/SimFL [Accessed on February 19th, 2024]
[30]https://github.com/EasyFL-AI/EasyFL [Accessed on February 19th, 2024]
[31]https://github.com/vivekkhimani/torchfl [Accessed on February 19th, 2024]
[32]https://github.com/APPFL/APPFL [Accessed on February 19th, 2024]
[33]https://github.com/NVIDIA/NVFlare [Accessed on February 19th, 2024]

|  | **FLEX** | PyS | TFF | FAT | Pad | Flo | IBM | Sub | OFL | FML | FJx | FLb | EFL | TFL | AFL | NVF |
|---|---|---|---|---|---|---|---|---|---|---|---|---|---|---|---|---|
| **Descentralized data sources:** | | | | | | | | | | | | | | | | |
| Built-in HFL datasets | 🟢 | 🔴 | 🟢 | 🔴 | 🔴 | 🟢 | 🔴 | 🔴 | 🔴 | 🔴 | 🟢 | 🟢 | 🟢 | 🔴 | 🔴 | 🔴 |
| Generate HFL data splits | 🟢 | 🔴 | 🔴 | 🔴 | 🔴 | 🟡 | 🔴 | 🔴 | 🔴 | 🟢 | 🔴 | 🔴 | 🟡 | 🔴 | 🔴 | 🔴 |
| Generate VFL data splits | 🟢 | 🔴 | 🔴 | 🔴 | 🔴 | 🔴 | 🔴 | 🔴 | 🔴 | 🔴 | 🔴 | 🔴 | 🔴 | 🔴 | 🔴 | 🔴 |
| Generate TFL data splits | 🟢 | 🔴 | 🔴 | 🔴 | 🔴 | 🔴 | 🔴 | 🔴 | 🔴 | 🔴 | 🔴 | 🔴 | 🔴 | 🔴 | 🔴 | 🔴 |
| **Distribution of FL nodes:** | | | | | | | | | | | | | | | | |
| Support for client-server architecture | 🟢 | 🟢 | 🟢 | 🟢 | 🟢 | 🟢 | 🟢 | 🟢 | 🟢 | 🟢 | 🟢 | 🟢 | 🟢 | 🟢 | 🟢 | 🟢 |
| Custom FL architectures | 🟢 | 🟡 | 🔴 | 🔴 | 🟢 | 🔴 | 🔴 | 🔴 | 🔴 | 🔴 | 🔴 | 🔴 | 🔴 | 🔴 | 🔴 | 🔴 |
| **Customized FL flows:** | | | | | | | | | | | | | | | | |
| Custom HFL flows | 🟢 | 🟢 | 🟢 | 🟢 | 🟢 | 🟢 | 🟢 | 🟢 | 🟢 | 🟢 | 🟢 | 🟢 | 🟢 | 🟢 | 🟢 | 🟢 |
| Custom VFL flows | 🟢 | 🟡 | 🔴 | 🟡 | 🟡 | 🔴 | 🔴 | 🔴 | 🟡 | 🟡 | 🔴 | 🔴 | 🔴 | 🔴 | 🔴 | 🔴 |
| Custom FTL flows | 🟡 | 🔴 | 🔴 | 🟡 | 🟡 | 🔴 | 🔴 | 🔴 | 🔴 | 🔴 | 🔴 | 🔴 | 🔴 | 🔴 | 🔴 | 🔴 |
| **Other properties:** | | | | | | | | | | | | | | | | |
| Open Source | 🟢 | 🟢 | 🟢 | 🟢 | 🟢 | 🟢 | 🟢 | 🟢 | 🟢 | 🟢 | 🟢 | 🟢 | 🟢 | 🟢 | 🟢 | 🟢 |
| Support for popular ML frameworks | 🟢 | 🟢 | 🟢 | 🟡 | 🟡 | 🟢 | 🟢 | 🟢 | 🟢 | 🟢 | 🟡 | 🟢 | 🟡 | 🟡 | 🟡 | 🟢 |
| Documentation and tutorials | 🟢 | 🟢 | 🟢 | 🟢 | 🟡 | 🟢 | 🟢 | 🟡 | 🟢 | 🟢 | 🟢 | 🟡 | 🔴 | 🟢 | 🟡 | 🟡 |

Table 1: Comparison of FL frameworks in terms of their FL simulation capabilities (🟢: full support; 🟡: partial support; 🔴: no support;).

Three degrees of compliance have been considered, depending on whether an aspect is supported by a framework or not. The green dots indicate that this aspect is fully supported by the framework. The orange ones indicate that the aspect is partially covered in the framework, i.e., it covers some cases but not all. Finally, the red dots mean that this aspect is not supported in the framework.

Clearly, most FL frameworks excel at simulating HFL, providing built-in HFL datasets and supporting popular ML frameworks. Moreover, they provide good enough documentation and tutorials. In contrast, these frameworks are limited to simulating client-server architectures and they lack support for VFL and TFL, at two levels, simulating data splits and supporting customized flows. Consequentially, these FL categorizations are not as popular among the research community as HFL and are vastly underrated, while they offer similar or improved privacy. The goal of FLEX is to boost research opportunities in the FL field by expanding its capabilities.

The development of frameworks like FLEX that target these specific challenges is a positive step in advancing FL research. By providing better support for complex FL architectures, VFL and TFL, you can potentially enable more research and innovation in these areas. As the field continues to evolve, addressing the limitations and gaps in existing frameworks will be essential to encourage broader adoption and exploration of FL in different domains. It's worth noting that the popularity of various FL categories can be driven by the practical use cases, the level of support from the research and development community, and the specific challenges that each category addresses. With more work and dedicated frameworks like FLEX, VFL and TFL might gain more recognition and adoption over time.

# 6 Concluding remarks and future work

FLEX represents a versatile and robust framework that empowers the FL community in multiple ways. It offers flexibility and a significant degree of power, enabling the development of FL versions for well-established ML fields. By addressing the existing challenges associated with simulating FL and filling the gaps in the predominantly HFL-focused landscape, FLEX strives to be a valuable resource for both researchers and developers in the FL community. Its primary goal is to facilitate experimentation and innovation while pushing the boundaries of FL capabilities.

Furthermore, FLEX is committed to enhancing FL support by providing libraries tailored for various ML domains. These libraries not only extend the capabilities of FLEX but also contribute to the continual improvement and expansion of the FL paradigm. In doing so, FLEX is poised to play a pivotal role in advancing the field of FL and addressing the evolving privacy and collaboration challenges in ML.

Future work primary focus will be on enhancing the simulation of FL and providing comprehensive support for novel FL workflows and data partitioning. Looking forward, our goal is to advance FLEX into a version that not only retains its inherent flexibility and feature richness but also implements genuine FL procedures. This advancement will allow users to seamlessly transition from a simulated FL application sketch within FLEX to a real-world FL scenario, bridging the gap between simulation and deployment. This evolution represents a crucial step towards the practical and efficient development of FL applications with FLEX.

# 7 Acknowledgments

This work is supported by the Recovery, Transformation and Resilience Plan, funded by the European Union (Next Generation Funds).